\documentclass[amsmath,aps,prl,twocolumn,superscriptaddress]{revtex4}

\usepackage{dcolumn}
\usepackage{bm}
\usepackage{amssymb}
\usepackage{color}
\usepackage[pdftex]{graphicx}

\usepackage{amsmath}
\usepackage[linesnumbered,ruled]{algorithm2e}

\usepackage{ulem}

\usepackage{url}
\usepackage[section]{placeins}
\usepackage[colorlinks=true,linkcolor=blue,citecolor=blue]{hyperref}

\begin{document}

\title{Ice, glass, and solid phases in artificial spin systems with quenched disorder}

\author{Yifei Shi} 
\affiliation{Department of Physics, University of Virginia, Charlottesville, VA 22904, USA}


\author{Cristiano Nisoli}
\affiliation{Theoretical Division and Center for Nonlinear Studies,
Los Alamos National Laboratory, Los Alamos, New Mexico 87545, USA}

\author{Gia-Wei Chern}
\affiliation{Department of Physics, University of Virginia, Charlottesville, VA 22904, USA}

\begin{abstract}
We present a numerical study on a  disordered artificial spin-ice system which interpolates between the long-range ordered square ice and the fully degenerate shakti ice.  Starting from the square-ice geometry, disorder is implemented by adding vertical/horizontal magnetic islands to the center of some randomly chosen square plaquettes of the array, at different densities. When no island is added we have ordered square ice. When all square  plaquettes have been modified we obtain shakti ice, which is disordered yet in a topological phase corresponding to the Rys F-model. In between, geometrical frustration due to these additional center spins disrupts the long-range Ising order of square-ice, giving rise to a spin-glass regime at low temperatures. The artificial spin system proposed in our work provides an experimental platform to study the interplay between quenched disorder and geometrical frustration.
\end{abstract}

\maketitle


Glass formation is a generic phenomenon that emerges in highly frustrated systems~\cite{bezard87,villain79,shintani06,tarjus05}.  For magnets with a well-defined order in the ground state, the presence of quenched disorder, either in the form of random interactions or vacancies, renders the pair-wise interaction frustrated. A glass transition occurs when the disorder is  strong enough to completely disrupt the coherent propagation of ordering through the system. Interestingly, when the frustration is ``maximized", as in geometrically frustrated magnets, instead of the glass phase, one obtains a liquid phase with extensively degenerate micro-states~\cite{moessner06,geo_frustration}. Adding quenched disorder to this degenerate manifold of frustrated magnets, the flat energy-surface associated with the huge degeneracy is expected to be replaced with a rugged energy landscape~\cite{yang15,cepas12,samarakoon16}. Spin-freezing occurs when temperature is lowered below the typical energy variation in the landscape. This mechanism has been proposed to explain glass transition in several frustrated magnets~\cite{saunders07,bilitewski17,shinaoka11}. Another scenario, first proposed for dilute pyrochlore spin ice, describes a glass transition that is induced by the {\rm frustrated} random interactions between emergent degrees of freedom in the degenerate manifold~\cite{sen15,sen12}.

In this paper, we propose an artificial spin-ice system to investigate the intricate competitions between magnetic ordering, geometrical frustration, quenched disorder, and spin-freezing. Artificial spin ices are meta-materials composed of nanoscale single-domain ferromagnetic elements arranged in certain geometries such that pairwise inter-island interactions are frustrated~\cite{wang06,nisoli13,chern21,marrows21}. Its namesake refers to the fact that quasi-degenerate magnetic states in the standard square-lattice geometry, which is a 2D version of the frustrated pyrochlore lattice, are characterized by the two-in-two-out local constraints~\cite{nisoli13,chern21}, similar to the short-range proton ordering in water ice~\cite{bernal33}. These systems have been the subject of extensive research for novel magnetic phases and dynamical phenomena.  In particular, the fact that collective modes that violate the local constraints carry a finite magnetic charge opens the possibility of a new technology, dubbed magnetricity, that utilizes the motion of such emergent magnetic charges, or monopoles, in devices~\cite{skjarvo20,cumings14,heyderman13}.


Because such nanoscale ferromagnets are produced lithographically, they can be arranged into essentially any 2D array pattern~\cite{gilbert16,morrison13,chern13,gilbert14,lao18,sklenar19,farmer16,brajuskovic16,shi18}. Moreover, using imaging techniques such as magnetic force microscopy and photoemission electron microscopy, the frustration-induced collective behaviors are both controllable and measurable at the level of a single magnetic moment. Artificial spin-ice systems have thus provided platforms for the study of a range of novel collective behaviors. Earlier experiments focused on athermal dynamics of artificial spin ices, partly because of the difficulty to reverse the magnetization of (relatively) large single-domain island with ambient thermal energies.  Various techniques have now been developed to fully thermalize the artificial spin systems~\cite{morgan11,zhang13,arnalds14,farhan13,kapaklis14}, thus opening the avenue to studying novel thermodynamic phases and phase transitions through specially engineered arrays.

\begin{figure}[t]
\includegraphics[width=0.99\columnwidth]{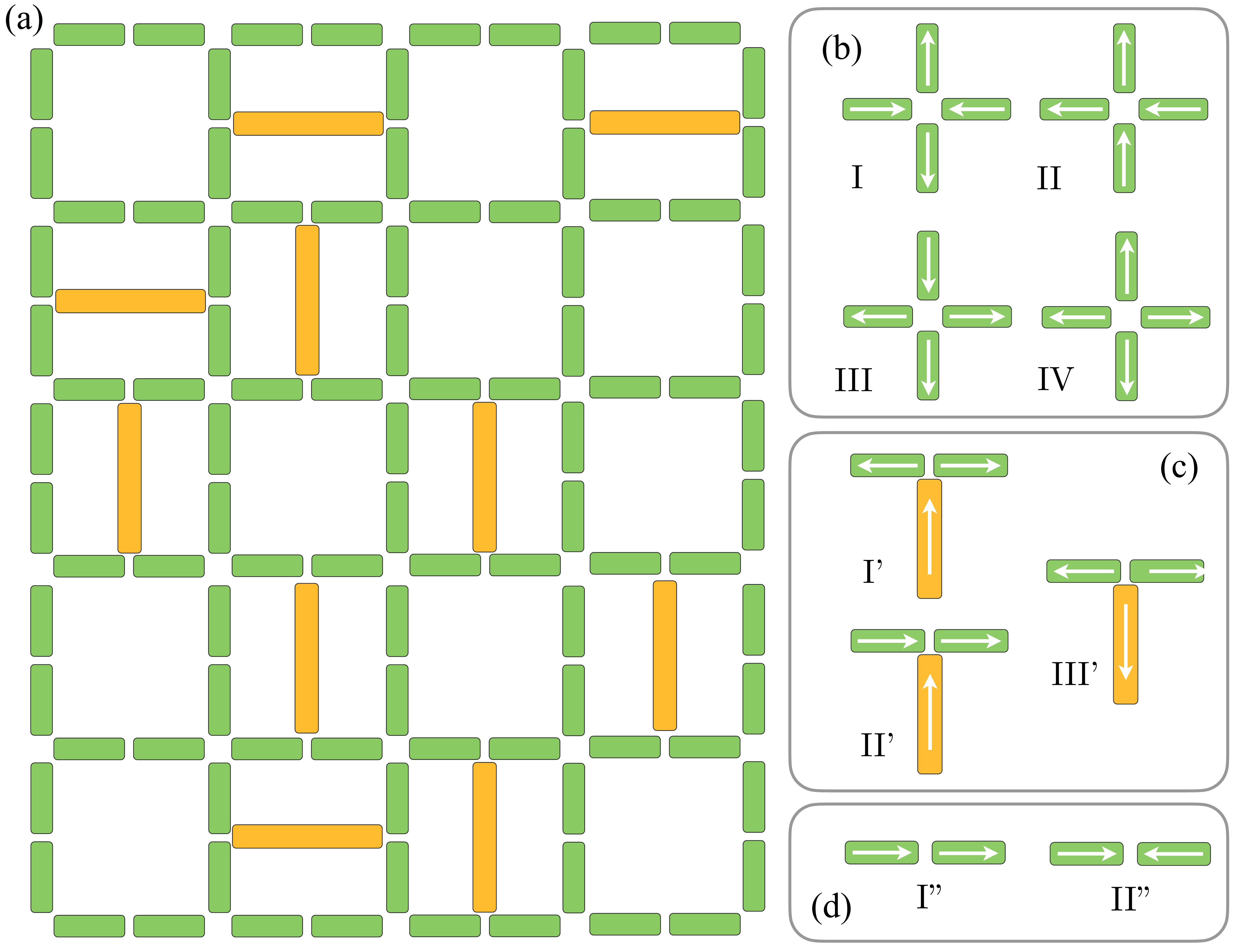}
\caption{(Color online) (a) Geometry of the disordered shakti spin ice. Panels~(b), (c), and (d) illustrate the four different types of four-legged ($z_4$) vertices,  three types of three-legged ($z_3$) vertices, and two types of two-legged ($z_2$) vertices, respectively. In this work, the energy of the various vertex-types are: $\epsilon_{\rm I} = -3.6$, $\epsilon_{\rm II} = -2$, $\epsilon_{\rm III} = 0$, and $\epsilon_{\rm IV} =  7.6$ for $z_4$ vertices, and $\epsilon_{\rm I'} = -1.8$, $\epsilon_{\rm II'} = -1$, $\epsilon_{\rm III'} = 3.8$ for $z_3$ vertices, and $\epsilon_{\rm I''} = 0$, $\epsilon_{\rm II''} = 8$ for $z_2$ vertices (in arbitrary energy unit). These energies are estimated based on realistic dumbbell models for nano-islands. }
\label{fig:lattice}
\end{figure}
 
Here, we consider an artificial spin system shown in Fig.~\ref{fig:lattice}(a), which can be viewed as an increasingly disordered version of the so-called shakti spin ice introduced in Ref.~\cite{morrison13,chern13,gilbert14}. We perform extensive Monte Carlo simulations to investigate effects of quenched disorder in this model and obtain its phase diagram
We will show that this special artificial spin-ice exhibits an emergent glass regime at low temperatures that is sandwiched by a long-range Ising order and the 2D version of the water-ice phase. 
Assuming nearest-neighbor island-island interactions, the energetics of the spin-ice system can be expressed in terms of vertex models~\cite{Lieb1972}, which is also the basis for our Monte Carlo simulations. 
The building blocks of this system, summarized  in Fig.~\ref{fig:lattice}(b)--(d),  include four types  of 4-legged vertices, three types of 3-legged vertices with orthogonal islands, and two distinct two-legged vertices. For convenience, we also call these three sets the $z_4$, $z_3$, and $z_2$ vertices, respectively. For each vertex-type, there are two spin configurations related to each other by time-reversal symmetry.  

\begin{figure}
\includegraphics[width=0.99\columnwidth]{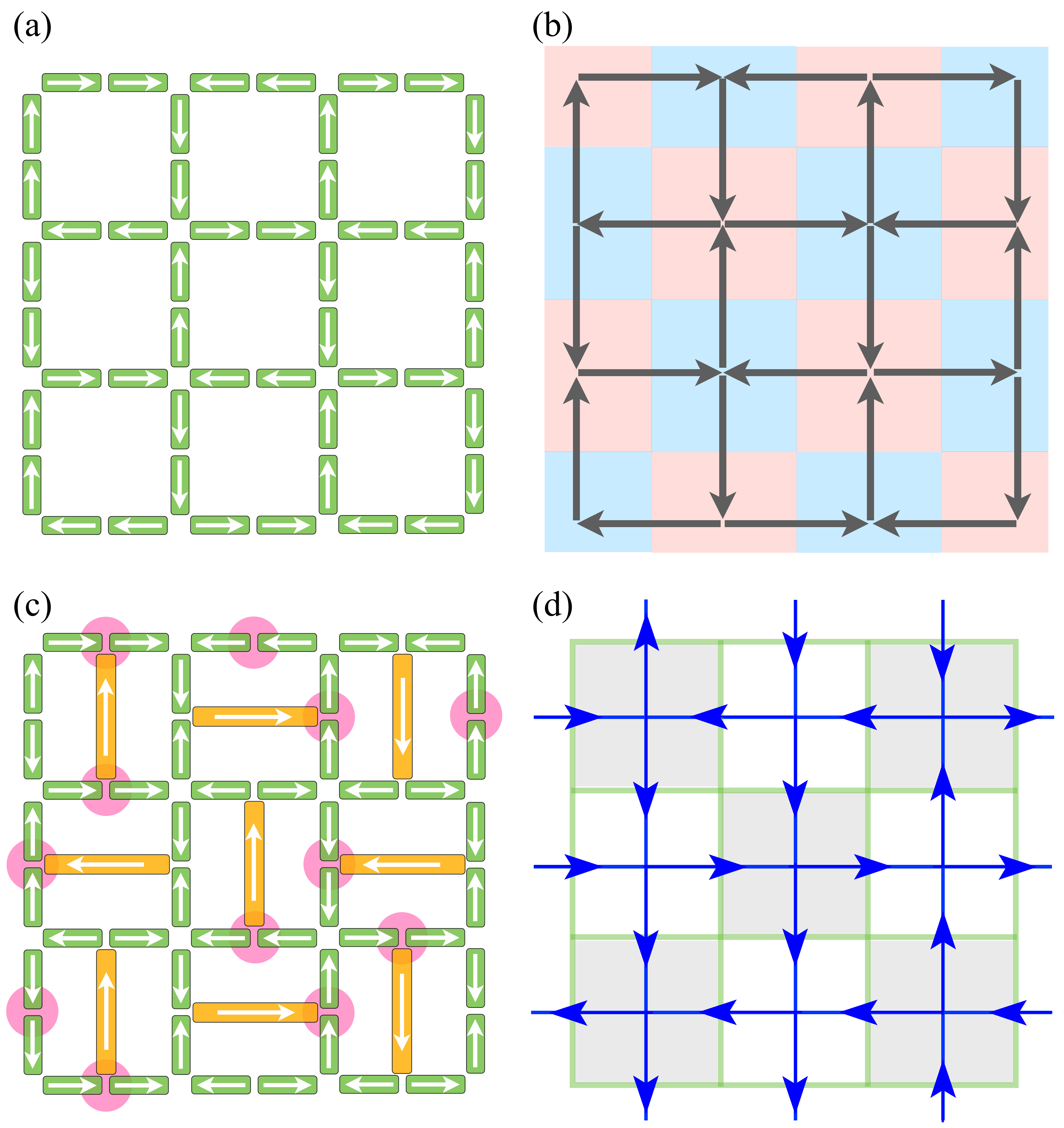}
\caption{(Color online) (a) The ground-state configuration in the square-ice limit with $f = 0$. The magnetization of the two nano-islands along any edge are aligned, while all $z_4$ vertices are of type-I. (b) shows the staggered arrangement of the $z_4$ vertices in the ground state; the red and blue  indicate the two type-I vertices related by time-reversal operation. (c) show one example of ground states in the shakti ice limit $f = 1$. The blue circle indicates the unhappy or frustrated type-II' $z_3$ vertices. The mapping to the corresponding arrow-representation in the six-vertex model is shown in panel~(d). The unhappy type-II' vertices are associated with an arrow pointing from even-sublattice plaquette (shaded) to odd-sublattice ones, thus realizing an emergent F-model at the critically disordered free fermion point~\cite{chern13}, where the ensemble is also equivalent to a dimer cover~\cite{lao18}.}
\label{fig:lattice2}
\end{figure}

The disorder of this artificial spin system is controlled by the fraction $f$ of square plaquettes which contain a central island. The $f = 0$ lattice without any central islands is simply equivalent to the widely studied square-ice shown in Fig.~\ref{fig:lattice2}(a). For convenience, using the square-ice as a reference, we also use the term {\it defect plaquette} to denote plaquette with a center island. The frustrated shakti ice, on the other hand, corresponds to the $f = 1$ limit and is obtained by alternatively placing an additional vertical or horizontal island in every square plaquette of the square ice; see~Fig.~\ref{fig:lattice2}(c). For $0 < f < 1$, the nano-array itself is disordered. Importantly, while the distribution of defect-plaquettes is completely random for a given fraction $f$, the orientation of the central island is not. By dividing plaquettes into even and odd sublattices, the inserted central islands must be vertical (horizontal) for even (odd) plaquettes. Following this rule, the structurally disordered lattice then smoothly connects to the shakti ice when $f = 1$.

\begin{figure}[t]
\includegraphics[width=0.99\columnwidth]{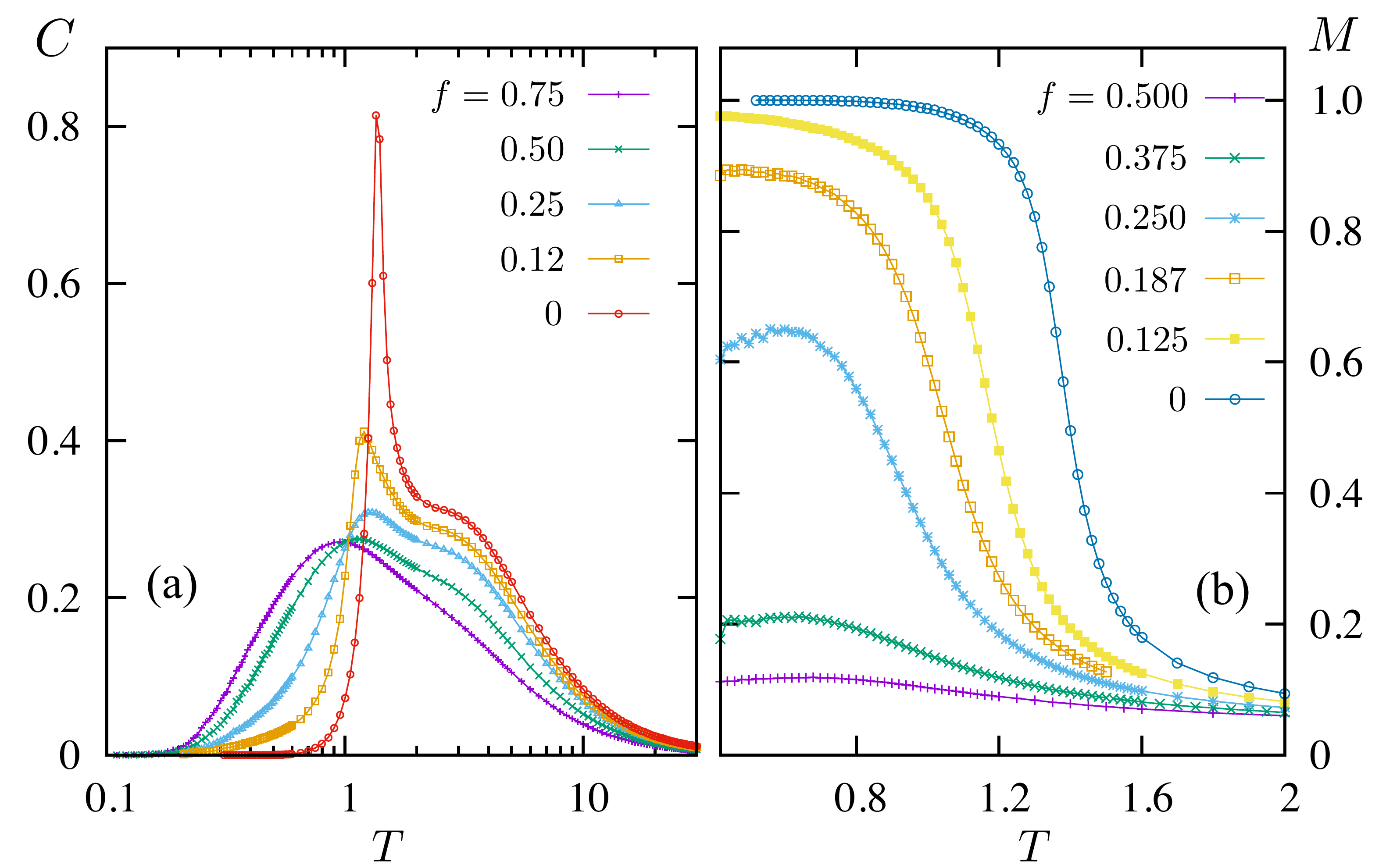}
\caption{(Color online) Monte Carlo simulations of disordered shakti ice of linear size $L = 16$, corresponding to $N_{\square} = L^2 = 256$ square plaquettes. (a) The evolution of specific versus temperature curves with increasing fraction $f$ of void plaquettes, i.e. squares without a central island. (b)~Normalized Ising order parameter $M$ as a function of temperature for varying fractions. The results for $f > 0$ are obtained by averaging over 50 different disorder realizations.}
\label{fig:ising-order}
\end{figure}

We note that an ``edge" of the $f = 0$ square-ice is composed of two magnetic islands forming a $z_2$ vertex. However, because of the huge energy cost for head-to-head or tail-to-tail $z_2$ vertices, the magnetization of the two islands in an edge align with each other at low temperatures, making the $f=0$ system equivalent to the standard square ice. As in the case of square-ice, the planar geometry breaks the degeneracy of the 2-in-2-out vertices into type-I and II, with the symmetric type-I arrangement having the lowest energy. As a result, the ground state in the $f=0$ square-ice limit is a staggered, antiferromagnetic arrangement of the type-I vertices, illustrated in Fig.~\ref{fig:lattice2}(a) and (b). The resultant long-range order is similar to the N\'eel order in antiferromagnetic Ising model~\cite{wang06,morgan11}, and is characterized by a $Z_2$-type Ising order. As shown in our Monte Carlo simulations in Fig.~\ref{fig:ising-order}, the transition to the ordered phase is marked by a pronounced peak in the specific heat versus temperature curve, below which the Ising order parameter quick rises to its maximum.

The shakti ice at the opposite $f = 1$ limit, on the other hand, is fully frustrated. Since every plaquette is now occupied by a center island, with alternating orientations, the magnetic nano-array is again free from disorder in the shakti ice limit. Here we assume that the nearest-neighbor island-island coupling is designed in such a way that the energy difference between type-I and II vertices is larger than that between the 3-legged type-I' and II'. Consequently, all four-legged vertices are in the lowest-energy type-I configuration, and frustration comes from the fact that some of the $z_3$ vertices have to be in the higher energy type-II' state.  A micro-state in this degenerate manifold can be described by specifying the location of the unhappy type-II' vertices, as shown in Fig.~\ref{fig:lattice2}(c). The positions of these type-II' defects are highly correlated in the degenerate manifold. As demonstrated in Ref.~\cite{chern13}, the local defect configurations satisfy constraints which are exactly equivalent to the Bernal-Fowler ice rules. The mapping is simple: each plaquette can be viewed as a water molecule H$_2$O, with the center of the plaquette being the oxygen and the type-II' defect being the hydrogen atom. 
In the ground states of the shakti array, the ice rules then dictate that each plaquette has exactly two defect type-II' vertices. By associating the frustrated vertices with an arrow pointing from even to odd-sublattice plaquettes, these ice-rule states can be further mapped to the an exactly solvable six-vertex  F-model~\cite{chern13}.




\begin{figure}
\includegraphics[width=0.99\columnwidth]{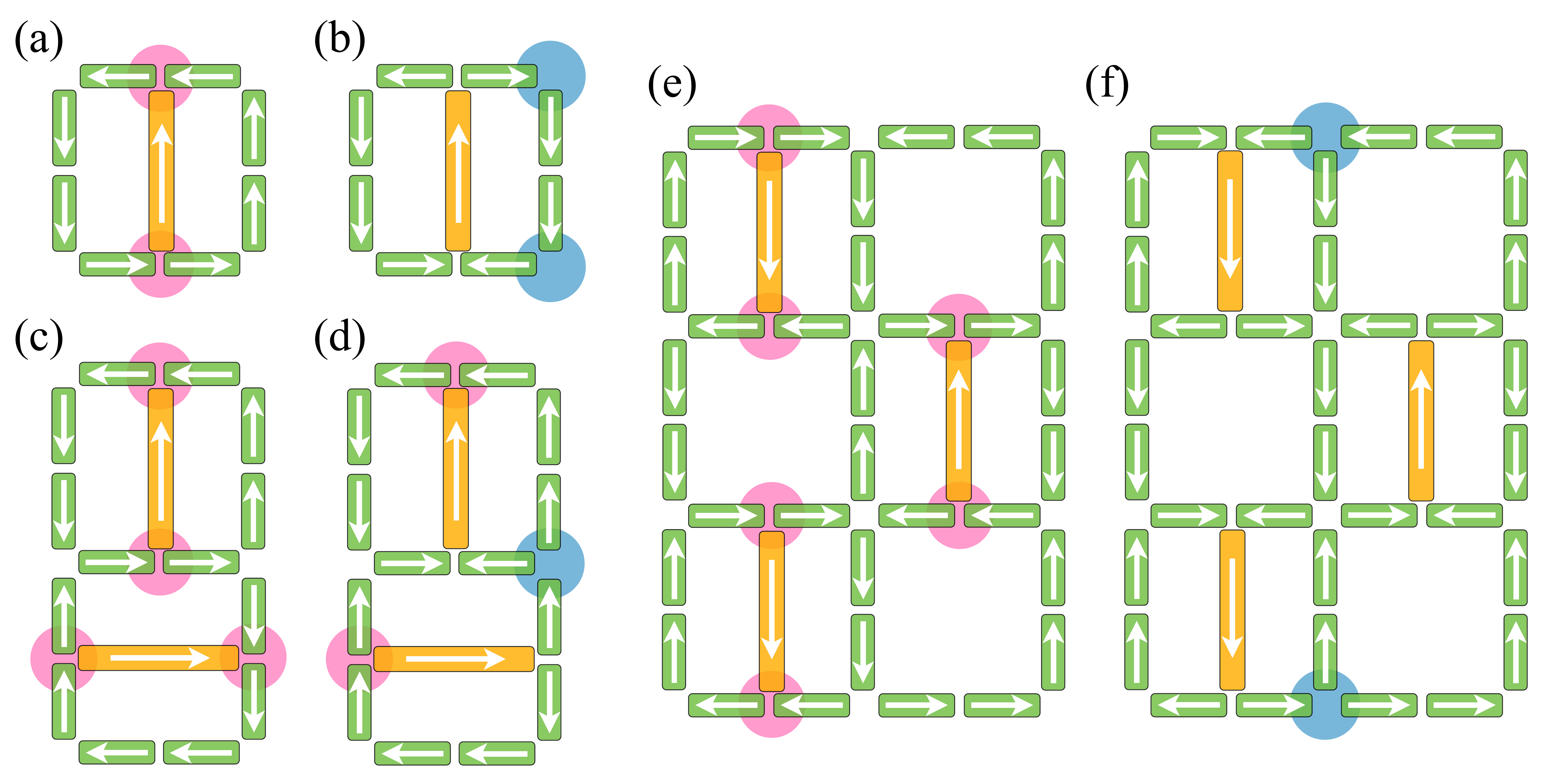}
\caption{(Color online) (a) and (b) show the two lowest-energy states of a plaquette with a center island in an otherwise perfectly Ising order of square ice. The configuration (a) in general is the ground state. Panels (c) and (d) show the two nearly degenerate configurations for two defect plaquettes next to each other. The lowest-energy states for 3 defect plaquettes are shown in panels~(e) and~(f). }
\label{fig:defect}
\end{figure}

\begin{figure}[t]
\includegraphics[width=0.98\columnwidth]{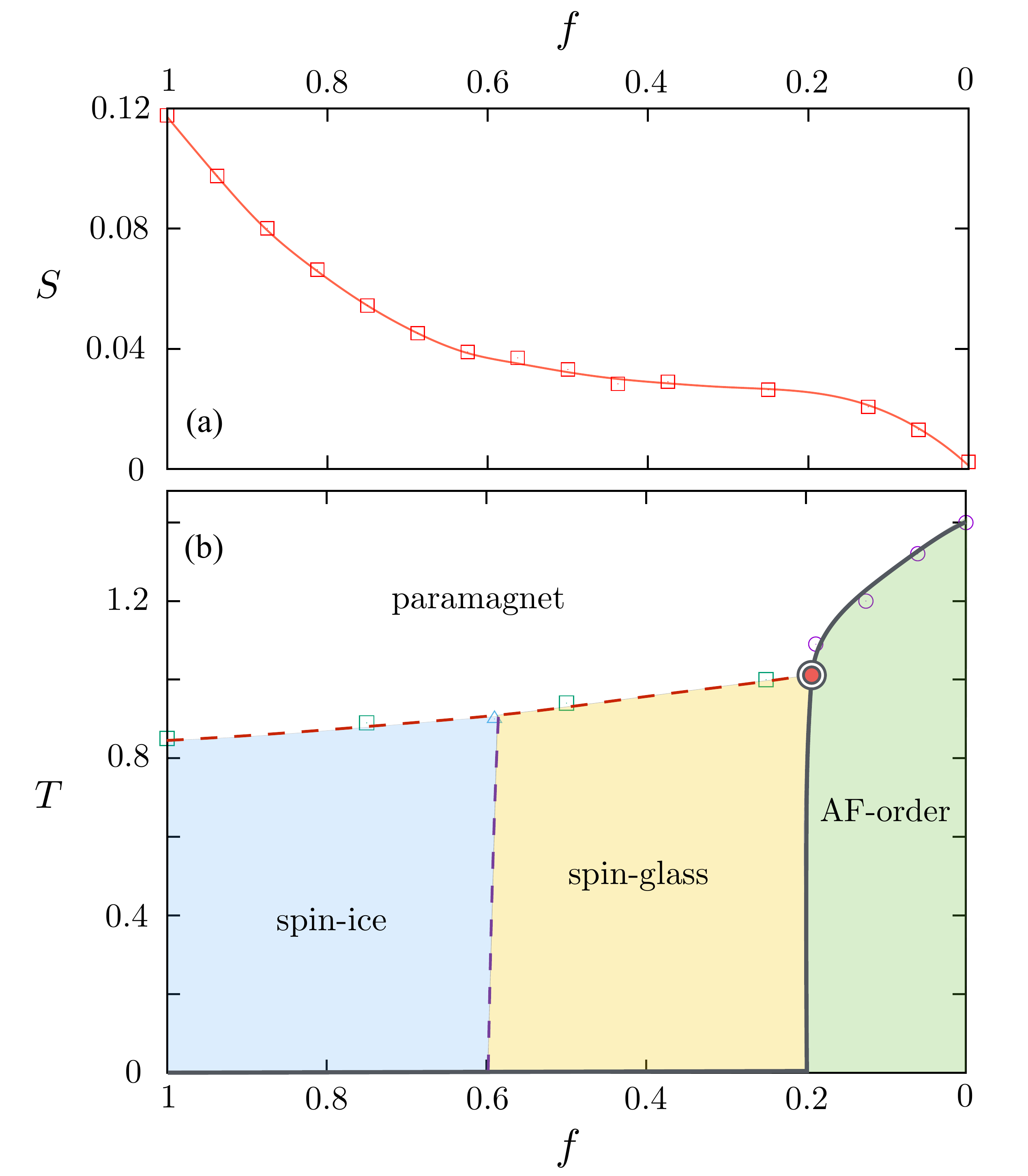}
\caption{(Color online) (a) The residual entropy $S$ obtained from specific-heat curves as a function of fraction $f$ obtained from Monte Carlo simulations on $L = 16$ systems. (b) Schematic phase diagram of disordered shakti spin ice in the $f$ versus $T$ plane. The dashed lines, indicating crossover from paramagnetic to low-$T$ disordered glass or ice phases, are estimated from peak of specific-heat curves. The cross-over line separating the glass and ice is estimated from the opposite finite-size behaviors of the spin-glass susceptibility. The solid line is a critical line because no symmetry breaking is possible in the glassy of paramagnetic phase. The red dot is generally called the Nishimori point~\cite{Nishimori1981}. The purple circles were estimated by finite-size scaling of the Ising transition. Since there is no ice or glass transition in two-dimension, the green squares corresponds to the maximum in the specific-heat peak.}
\label{fig:phases}
\end{figure}

Next we consider the intermediate regime $0 < f < 1$ with randomly distributed defect plaquettes. First, we examine the energetics of one such plaquette. Naively, one would expect the mixed configuration, namely a type-I' vertex and an unhappy type-II' vertex at the two ends of the center island, would be the lowest energy state, similar to those in the ground states of a shakti ice.  However, at low concentration $f$, the presence of a 3-legged type-II' vertex disrupts the Ising order of the square ice, and thus a costly type-III or multiple type-II vertices, will be created. The lowest-energy configuration turns out to be the one with two type-II' $z_3$ vertices as shown in Fig.~\ref{fig:defect}(a), while the first excited state has two type-II $z_4$ vertices shown in Fig.~\ref{fig:defect}(b). This also indicates that the presence of a few such defect plaquettes will not disrupt the long range order, which is also confirmed in our simulations; see the $f = 0.125$ curve in Fig.~\ref{fig:ising-order}(b). Moreover, with both ends being in the unhappy type-II' vertex, the center spin in Fig.~\ref{fig:defect}(a) can point in either up or down directions with equal energy. This double degeneracy thus gives rise to a residual entropy which is proportional to the number of central islands, i.e. $S \sim k_B f \ln2$ for small $f$, a result also confirmed in our simulations shown in Fig.~\ref{fig:phases}(a).

With higher density of defect plaquettes, new localized vertex-configurations which can disrupt the N\'eel ordering are stabilized energetically. For example, when two defect-plaquettes are next to each other as shown in Fig.~\ref{fig:defect}(c) and (d), annihilation of two type-II' $z_3$ vertices associated with the center islands produces a type-II $z_4$ vertex with nearly the same energy. Another example is shown in Fig.~\ref{fig:defect}(e) and (f), in which rearrangement of a few local spins transforms the six type-II' vertices into two type-II $z_4$ vertices, which lowers the energy. Importantly, it is known that type-II  vertices align themselves into domain walls separating opposite $Z_2$ orientations of the antiferromagnetic ground state~\cite{Nisoli2020}. Therefore, adjecent defect plaquettes, as depicted in Fig.~\ref{fig:defect}(d)-(f), must pin domain walls. Their presence  weakens the symmetry breaking of the staggered pattern of type-I vertices, thus reducing the Ising order of the system.  Indeed, as  shown in Fig.~\ref{fig:ising-order}, we find that the Ising order parameter is suppressed with increasing fraction $f$, while the corresponding peak in specific heat also transforms into a broad bump. The Ising order is estimated to disappear around $f = 0.25$, which is smaller than either the site or bond percolation thresholds.

The disappearance of N\'eel order with increasing $f$ is obviously attributable to the enhanced frustration induced by defect plaquettes. The system is expected to undergo a phase transition into a disordered regime. However, it is prima facie not obvious  whether the new phase is a glassy state or not, since we also know that the system at low $T$ enters an ice phase at large enough $f$. To investigate the nature of the disordered phase at intermediate $f$, we compute the Edward-Anderson order parameter $q_{\rm EA} =   \sum_i \sigma^{(a)}_i \sigma^{(b)}_i/N $ where the superscripts (a) and (b) denote two replicas of the system, obtained from identical disorder realization~\cite{edwards75}. A large $q_{\rm EA}$ indicates significant overlaps between different replicas of the system, thus pointing to a unique lowest energy state with disordered spins. Next, we compute the spin-glass susceptibility $\chi_{\rm SG} = N [ \langle q^2_{\rm EA} \rangle ]$, where $\langle \cdots \rangle$ means Monte Carlo average from independent runs, and $[ \cdots]$ indicates averaging over different disorder configurations. Parallel tempering~\cite{hukushima96} is used to improve thermalization of the disordered system.

\begin{figure}[t]
\includegraphics[width=0.99\columnwidth]{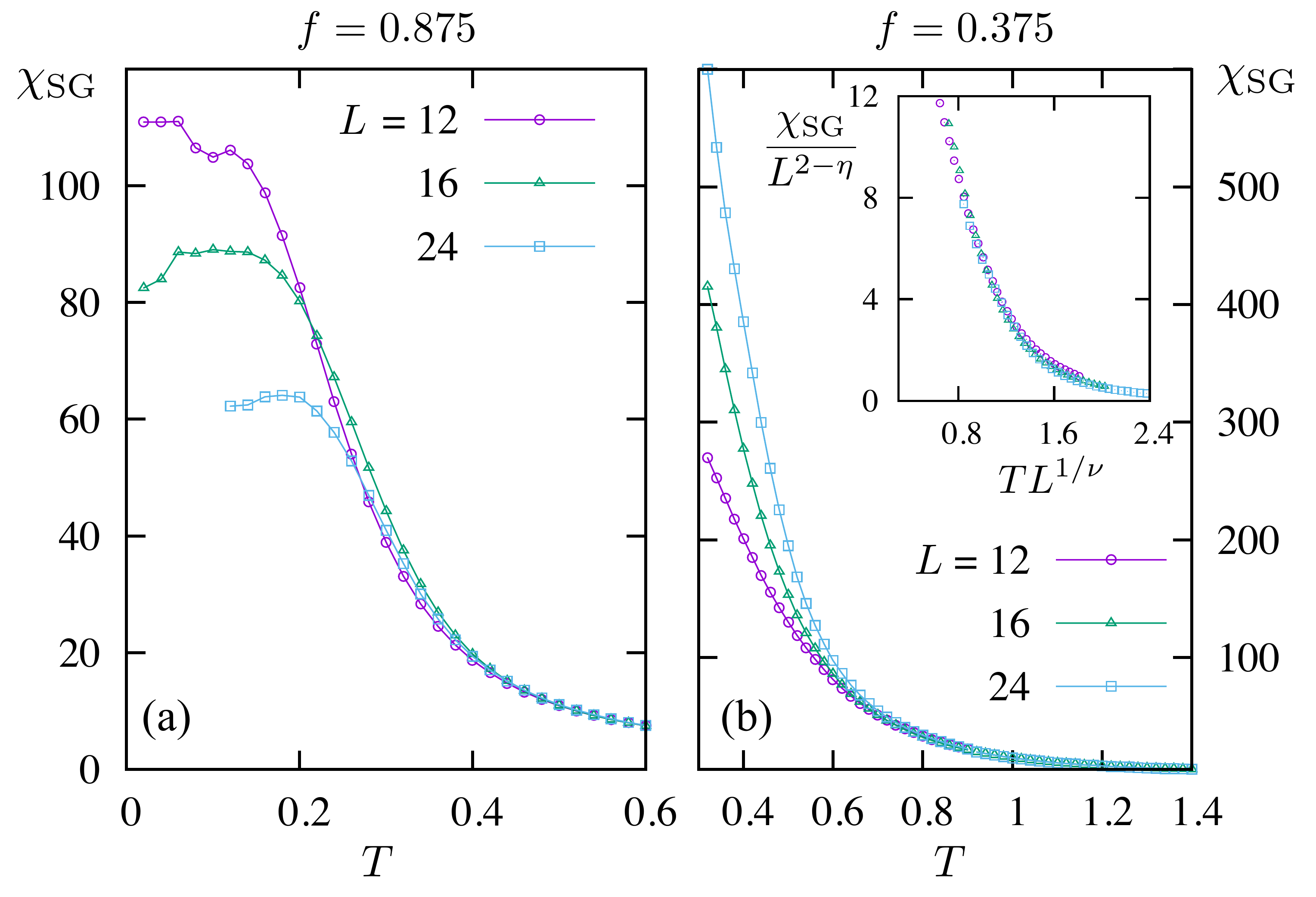}
\caption{(Color online) Spin glass susceptibility $\chi_{\rm SG}$ for fraction of defect plaquetts at (a) $f=0.125$ and (b) $f=0.625$. The inset in panel (b) shows finite-size scaling of susceptibility. The exponents $\eta = 0.2$ and $1/\nu = 0.38$ are used.   }
\label{fig:glass-order}
\end{figure}

Fig.~\ref{fig:glass-order} shows the spin-glass susceptibility for two different fractions $f$, obtained from Monte Carlo simulations of various system sizes. While the susceptibility  increases upon lowering the temperature, we find that $\chi_{\rm SG}$ reaches a maximum at low temperatures for large $f$, i.e. systems close to the shakti-ice limit. Also importantly, as shown in Fig.~\ref{fig:glass-order}(a), the susceptibility declines as the system size $L$ is increased, indicating that the spin-glass order might vanish in the thermodynamic limit. These results suggest that the six-vertex ice regime survives into a finite regime of~$f \lesssim 1$.

On the other hand, for intermediate fraction of defect plaquettes, e.g. $f = 0.375$, the spin-glass susceptibility seems to diverge when approaching zero temperature; see Fig.~\ref{fig:glass-order}(b). Also importantly, here we see a finite-size behavior completely opposite to that of large $f$ systems. The susceptibility is enhanced with increasing system sizes. Since there is no finite-temperature glass transition in 2D, we perform a finite-size scaling analysis by plotting $\chi_{\rm SG}/ L^{2 - \eta}$ versus rescaled temperature $T L^{1/\nu}$. This scaling relation is consistent with a conventional glass transition with a critical temperature at $T_g = 0$~\cite{morgenstern79,mcmillan83,singh86,bhatt88}. As shown in the inset of Fig.~\ref{fig:glass-order}(b), agreeable collapsing of data-points from different sizes is obtained by choosing exponents $\eta = 0.2$ and $1/\nu = 0.38$. Our result is consistent with the critical exponents of 2D Ising glass with random binary $\pm J$ bonds~\cite{bhatt88}. In particular, since there is a nonzero residual entropy in the glassy regime of our model, a nonzero $\eta$ here is consistent with transition to a glass ground state that possesses a finite degeneracy.

Combining our numerical results, we show in Fig.~\ref{fig:phases}(b) a schematic phase diagram in the $f$ versus $T$ plane for the disordered shakti spin-ice system. Also shown in panel~(a) is the residual entropy $S$, obtained by integrating the specific-heat curve, as a function of the fraction. For small density of defect-plaquettes, the system enters a Ising-ordered phase at low temperatures. The long-range N\'eel type Ising order is destroyed upon increasing the density of defect-plaquettes.  The resultant spin-disordered regime at low-temperature is characterized by a spin-glass susceptibility that diverges as $T$ tends to zero. A finite entropy remains in this spin-glass-like phase. As we further increase the number of defect-plaquettes, the residual entropy starts to increase significantly, signaling a crossover to a spin-ice phase, which becomes the exactly solvable ice phase in the well-known six-vertex F-model when $f = 1$. It is worth noting that, dynamically, there is perhaps no sharp distinction between the glass and ice phases, especially as far as the discrete local spin-update (either Metropolis or Glauber) is concerned. In our opinion, the main difference here is that the zero modes in the disordered ice phase ($f \lesssim 1$) are able to produce a macroscopic number of ground states that are significantly different from each other, while this is not the case for the spin-glass phase. 

The disordered shakti spin ice system provides a unique platform to experimentally study the collective behaviors and dynamical phenomena in both glass and ice regimes at the microscopic level. One intriguing topic is how the quenched structural disorder in this system affect the dynamical behaviors of type-III vertices, which behave as emergent magnetic monopoles. It is also interesting to see whether the scenario of topological spin-glass via ``ghost spins",  first proposed in diluted pyrochlore spin ice~\cite{sen15}, is realized in this artificial ice system with  quenched disorder.  We hope our work will motivate further experimental studies along these directions.


\bigskip

{\bf Acknowledgement}. We thank insightful discussions with Peter Schiffer and Ian Gilbert. CN's work was carried out under the auspices of the U.S. DoE through the Los Alamos National Laboratory, operated by Triad National Security, LLC (Contract No. 892333218NCA000001) and founded by DOE-LDRD. The authors also acknowledge the support of Advanced Research Computing Services at the University of Virginia.

\bigskip

{\bf Data Availability.} The data that support the findings of this study are available from the corresponding author upon reasonable request.

\end{document}